\documentclass{JHEP3}
\usepackage{epsfig}
\usepackage{amsmath}
\usepackage{bbm}
\usepackage{cite}

\newcommand{\bee}{\begin{equation}}
\newcommand{\ee}{\end{equation}}
\newcommand{\beea}{\begin{eqnarray}}
\newcommand{\eea}{\end{eqnarray}}

\def\re{\mathrm{Re}}

\title{More about QCD on compact spaces}

\author{Thomas DeGrand$^a$, Roland Hoffmann$^{ab}$, Johannes Najjar$^{ac}$ \\
\it $^a$Department of Physics, University of Colorado, Boulder, CO 80309 USA \\
\it $^b$Department of Physics, University of Wuppertal, 42097 Wuppertal, Germany \\
\it $^c$Institute for Theoretical Physics, University of Regensburg, 93040 Regensburg, Germany \\\\
  E-mail: \email{degrand@aurinko.colorado.edu}\\
  \phantom{E-mail: }\email{hoffmann@pizero.colorado.edu}\\
	\phantom{E-mail: }\email{johannes.najjar@physik.uni-regensburg.de}
}

\date{\today}
\abstract{
We present some results about spontaneous breaking of global symmetries
for four-flavor, three color QCD on compact spaces with two short directions.
When the two short directions have equal length and identical boundary conditions, there
 is a single transition.
When the two short directions have boundary conditions of opposite parity and
are of roughly equal extent, the C-breaking and
 deconfinement transitions separate. When the two short dimensions are of different length,
the transitions are modified in qualitative agreement with expectations from dimensional reduction.
These features resemble the situation in pure gauge simulations at small and large number of colors.}

\begin{document}

\section{Introduction}
The geometry of an embedding space can influence the phase structure of a field theory.
The most familiar example of such behavior is the use of a compact temporal dimension to
study a field theory at finite temperature: when the compact dimension is sufficiently
small, the theory can undergo a phase transition. If the field theory is a gauge theory,
the transition is typically a passage from a confined phase to a deconfined one, and the 
order parameter is the Polyakov line wrapping around the shortest dimension.
For an $SU(N)$ gauge theory, the Polyakov loop orders along one of the elements
of the $Z(N)$ center of the gauge group.

When the compact space is used to describe thermodynamics, fermion fields have
antiperiodic boundary conditions in the temporal/thermal direction. Fundamental
representation fermions break the $Z(N)$ center symmetry in the action, but
a phase transition might still occur. When it does, the Polyakov loop takes an
expectation value which is real; the other $Z(N)$ orientations are disfavored.
This is not the case if the short direction in one in which fermions have periodic
boundary conditions.
Last year \"Unsal and Yaffe\cite{Unsal:2006pj} (building on
earlier work by them and by Kovtun \cite{Kovtun:2003hr,Kovtun:2004bz,Kovtun:2005kh})
pointed out that the effects of fermions on these transitions is strongly dependent
on their boundary conditions, and that when fermions had periodic
boundary conditions in the shortest dimension, the most likely possibility is that the fermions drive 
the system into a phase of broken charge conjugation (basically by forcing the
Polyakov loop into one of the complex directions of $Z(N)$).
This ordering behavior was anticipated almost twenty years ago by 
van Baal\cite{vanBaal:1988va,vanBaal:2000zc}. 

Two of us\cite{DeGrand:2006qb} recently performed simulations in
$SU(3)$ with fundamental-representation fermions, which revealed this behavior.
The only related numerical work we are aware of is by Lucini, Patella, and Pica\cite{Lucini:2007as},
who associated a persistent baryon current around the compact direction with the breaking
of center symmetry.

In this note we extend our previous work and explore what happens when
there are two small compact dimensions. We expect (and see) that when the two short directions
have identical (periodic) boundary conditions, there is still a single critical point.
However, when one of the directions is periodic and the other is antiperiodic,
there are separate ordering transitions for the two directions.

We then look briefly at the case where the two short directions have different lengths. We find that
there is still a symmetry breaking transition which orders the shortest length,
but that any transition in the next-shortest length is either washed out or pushed
to much smaller coupling (much higher bare $\beta$) than what it would have been if it were
the shortest direction.
It happens that similar behavior has been observed in simulations of pure gauge
theory: $SU(2)$ gauge theory in four and five dimensions by Ref. \cite{Farakos:2002zb},
and in simulations of large-$N$ gauge theory in three dimensions, by Refs. \cite{Bursa:2005tk}
and \cite{Narayanan:2007ug}.
We believe that one can make a qualitative explanation of our observations using
dimensional reduction.

Our simulations are completely standard:
we have three  dimensions with periodic boundary
conditions and one time dimension with 
anti--periodic boundary conditions for the
fermions. The gauge fields are periodic in all directions.

The present study uses unrooted (i.e. four--taste) staggered
fermions. We work with an improved action to minimize
cutoff effects.
We employ the Hybrid Monte Carlo algorithm.
Our code is based on the publicly available MILC
package\footnote{\texttt{http://www.physics.utah.edu/\~{}detar/milc/}}
for improved staggered quarks \cite{Orginos:1999cr,Bernard:2001av,Aubin:2004wf}
on a Symanzik gauge background.
We have taken two values for the quark mass, $am_q=0.05$ and 0.2. Our simulations
 typically use about 1200  molecular dynamics trajectories per point, with
significantly more for runs near the transition in difficult cases.

We use the phase of the Polyakov loop (which does not require renormalization)
as our order parameter.
We map the phase range between two $SU(3)$ center elements to the full 
circle by
taking $(P/|P|)^3$ and then project onto the real axis,
\bee
S(P)=\re (P/|P|)^3=\cos(3\arg P)\;. \label{defS}
\ee
When the system is in its unbroken phase, we expect that $\langle S(P)\rangle =0$;
when the Polyakov loop only takes values in $Z(3)$, $\arg P = 2\pi j/3$, $j=0,1,2$,
and $\langle S(P)\rangle =1$. We can define the location of a transition by
$\langle S(P)\rangle=1/2$.
To make this determination, we fit $S(P)$ to the phenomenological formula
\bee
S(P)_{\beta}=\frac{1}{2} (1 + \tanh(\alpha(\beta - \beta_{crit})))
\ee
where $\alpha$ is just an arbitrary parameter, while varying the number of $\beta$ values
we keep near the inflection point.

In the next section we give an overview of our numerical results, then we describe
the case of two equal-length short dimensions.
In the following section we consider short directions of different length and describe 
simulations of pure gauge theory which produce similar behavior.

\TABULAR[!t]{|l|lllll|}{
\hline
geometry & $4\!\times\! 10^2\!\times\! 10$ & $4\!\times\! 10^2\!\times\! 6$
 & $4\!\times\! 10^2\!\times\! 4$ & $6\!\times\! 10^2\!\times\! 4$
 & $10\!\times\! 10^2\!\times\! 4$ \\ 
\hline
$\beta_{crit}(x)$, $am_q\!=\!0.2$ & 6.28(1) & 6.20(1) & 6.46(1) & - & - \\ 
$\beta_{crit}(t)$, $am_q\!=\!0.2$ & - & - & 6.037(4) & 5.99(1) & 5.96(5)\\
\hline
$\beta_{crit}(x)$, $am_q\!=\!0.05$ & 5.750(6) & 5.760(7) & 6.12(2) & - & - \\ 
$\beta_{crit}(t)$, $am_q\!=\!0.05$ & - & - & 5.496(5) & 5.460(4) & 5.405(5)\\
\hline
}
{The critical $\beta$'s for $am_q=0.2$ and 0.05 for various geometries
 ($Nx\times Ny \times Nz \times Nt$).
 A dash indicates that we could not observe a transition.
\label{tab}}

\section{The thermal and C-breaking transitions}

\subsection{Two periodic directions}

We first performed simulations with two short directions of the lattice, both with periodic
boundary conditions for the fermions. On symmetry grounds, we expect to
see a single phase transition separating a confining phase from a phase 
where charge conjugation is broken. This phase is characterized by a Polyakov loop
oriented in one of the complex directions, in either or both of the two short directions.
To check this, we performed simulations at $am_q=0.2$ on a $4^2\times 10^2$ lattice.
 A graph of the relevant $S(P)$'s is shown
 in Fig.~\ref{fig:4x4x10x10m0.2xybeta}.
We observed a single critical point for a transition in the $x$ and $y$ directions,
and were unable to identify any correlations between the value of the Polyakov loops in the
two directions. Fitting the behavior of the two Polyakov loops separately
gave  $\beta_{crit}(x)=6.33(1)$ 
and $\beta_{crit}(y)=6.31(1)$, which is consistent within uncertainties of the presence of
a single transition. The transition is pushed to somewhat weaker coupling than
$\beta_{crit}$ for a $4\times 10^3$ lattice (see Table \ref{tab}).
The Polyakov loops in the two directions do slightly communicate with each other.

\EPSFIGURE[!b]
{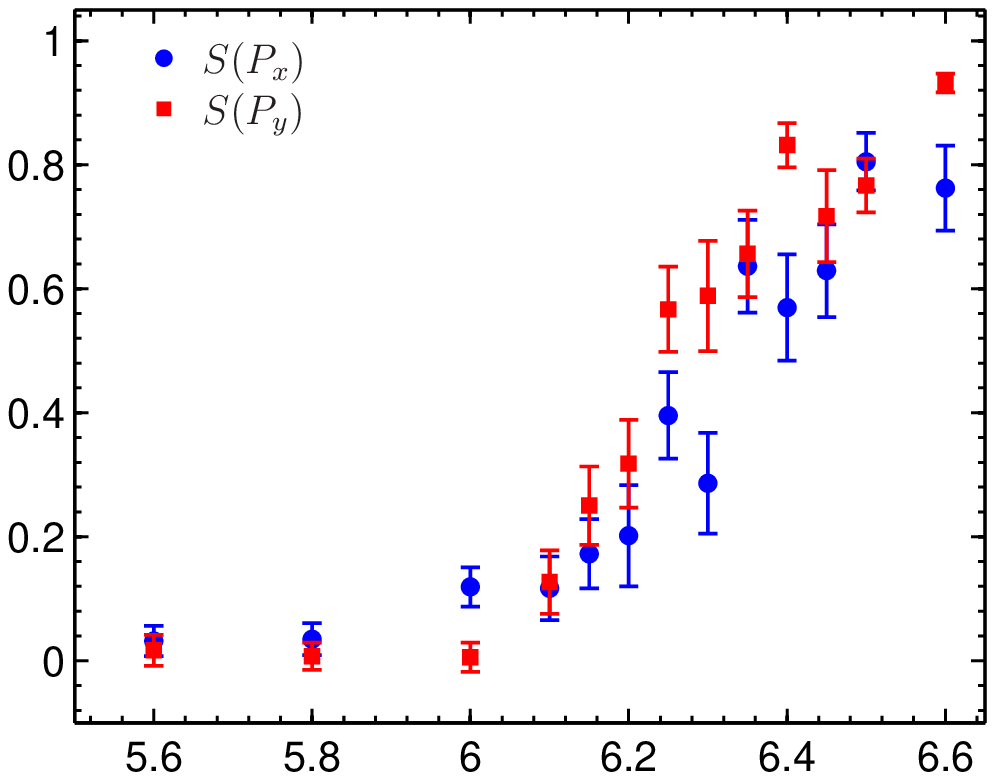, width=.6\textwidth}
{$S(P)$ in the $x$ and $y$ direction vs.
$\beta$ at $am_q=0.2$ for $4^2\!\times\! 10^2$ lattices.\label{fig:4x4x10x10m0.2xybeta}}

\subsection{One periodic and one antiperiodic direction}

We next turn to simulations in which the temporal direction and one of the spatial directions
have the same length, and both are much shorter than the other two directions.
In this case, since no symmetry relates the two short directions, we expect to see two separated
phase transitions, one in which the temporal Polyakov loop orders in 
a real direction and one in which
the spatial loop orients in a complex direction. Our results are 
shown in Table \ref{tab}.
An example of our observations is shown in Figs. 
\ref{fig:4x10x4m0.05tbeta}-\ref{fig:4x10x4m0.05xbeta}.

The transitions separate, with the $t$ transition, which occurs at
 lower $\beta$ remaining close to its value
 from simulations with only one short direction. The $x$ transition notices that the
Polyakov loop in the $t$ direction has ordered and shifts to higher $\beta$ than its
value when the $t$ direction is long and the Polyakov loop is disordered.

\section{QCD in asymmetric spaces}

\subsection{Observations}
We now consider the case that we have two short directions of different size,
one with periodic and one with antiperiodic boundary conditions for the fermions.
In particular, we take $N=4$
for the shortest direction and $N=6$ for the next-shortest one.
Results are again summarized in Table \ref{tab}.
We found that
the Polyakov loop in the shortest direction continues to undergo an ordering transition
exactly as if there was only one short direction: that is, along the real axis if the shortest
direction had antiperiodic boundary conditions, or into one of the complex directions if the
boundary conditions were periodic. Its location shifts by a small amount.
However, once the shortest direction had ordered, the transition in the next-shortest direction
becomes very smooth (we cannot say if it has disappeared or not) and moves to very large $\beta$.
We illustrate this phenomena in Figs. \ref{fig:4x10x6m0.2tbeta} and \ref{fig:4x10x6m0.2xbeta},
from simulations with $am_q=0.2$.
Here the length in the periodic (``$x$'') direction is $L=4$ and the 
antiperiodic  (``$t$'') direction has $L=6$, so the transition in the $x$ direction persists
while the $t$ transition is lost.

\DOUBLEFIGURE[t]
{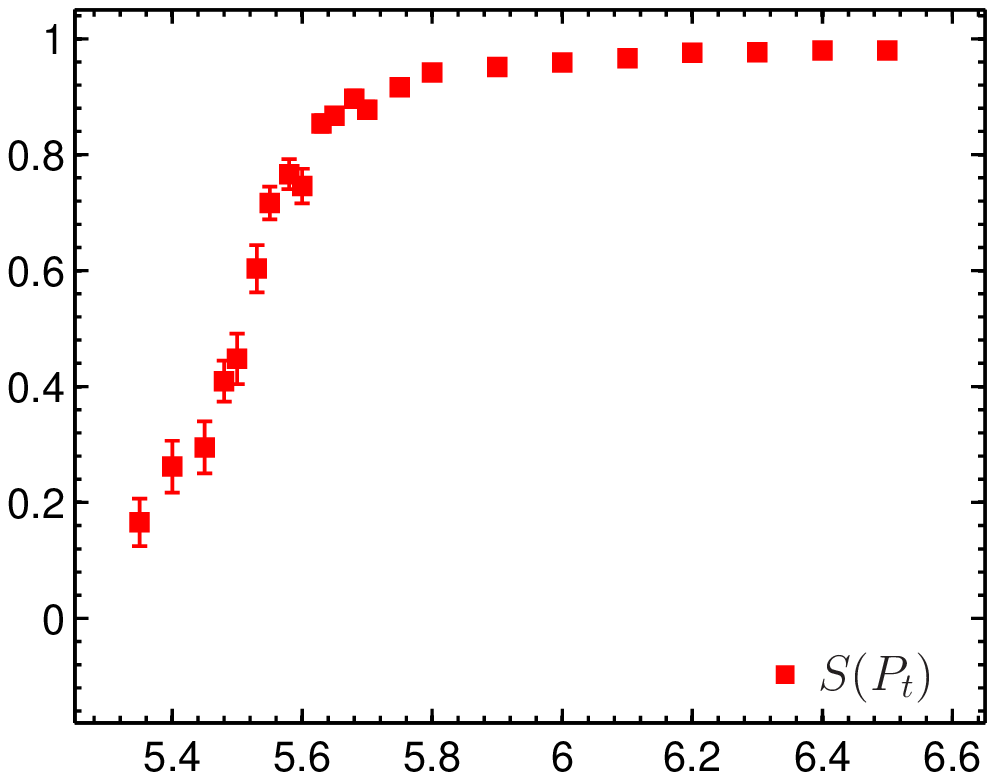, width=.47\textwidth}
{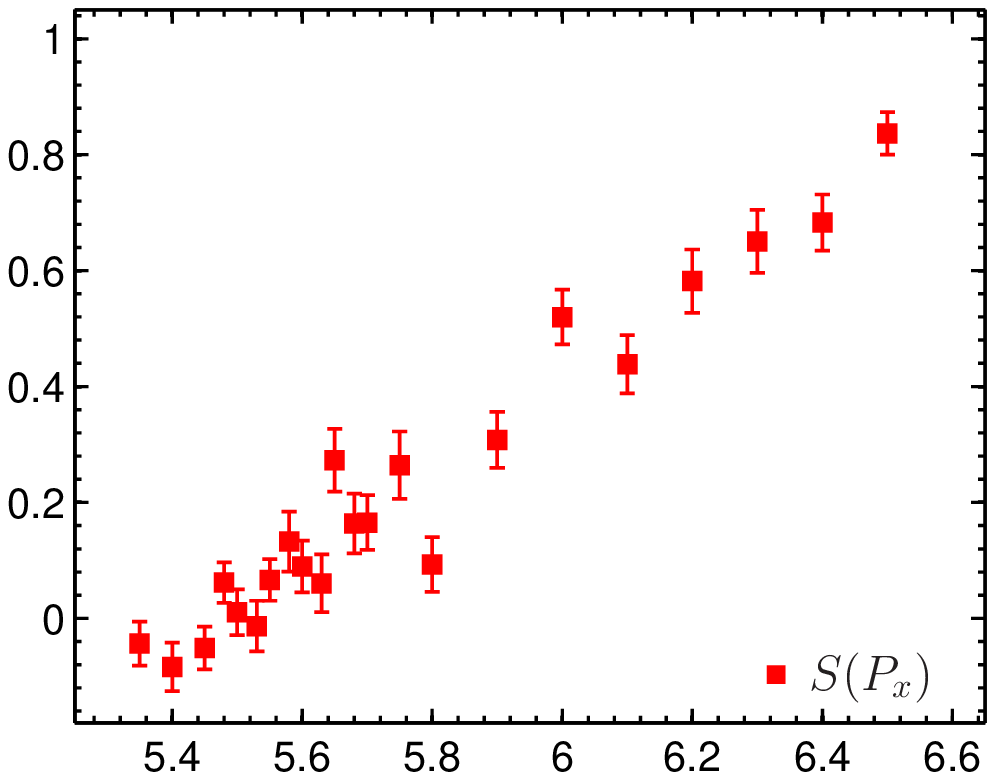, width=.47\textwidth}
{$S(P)$ in the $t$ direction vs. $\beta$ at $am_q=0.05$ for $4\!\times\! 10^2\!\times\! 4$ lattices.
\label{fig:4x10x4m0.05tbeta}}
{$S(P)$ in the $x$ direction vs. $\beta$ at $am_q=0.05$ for $4\!\times\! 10^2\!\times\! 4$ lattices.
\label{fig:4x10x4m0.05xbeta}}

When we make the $t$ direction shorter, the situation is reversed: the
 $t$ transition remains, while the
$x$ transition becomes very round and moves to very high $\beta$ or disappears.
Compare Figs. \ref{fig:6x10x4m0.2tbeta}-\ref{fig:6x10x4m0.2xbeta}.

\DOUBLEFIGURE[t]
{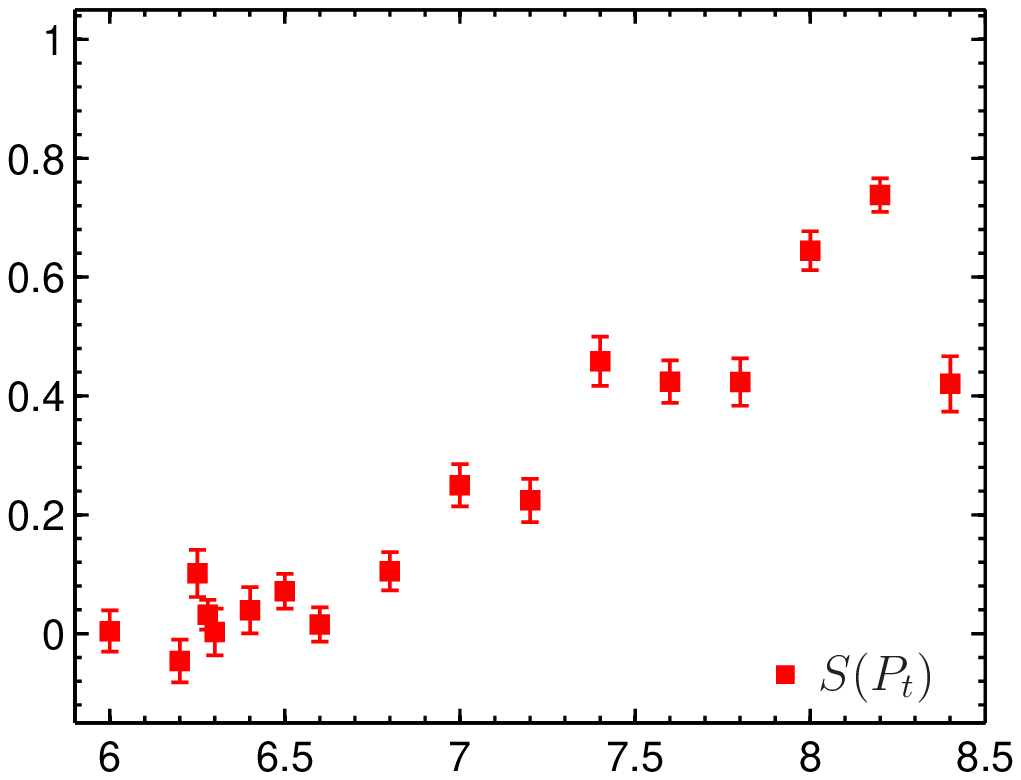, width=.47\textwidth}
{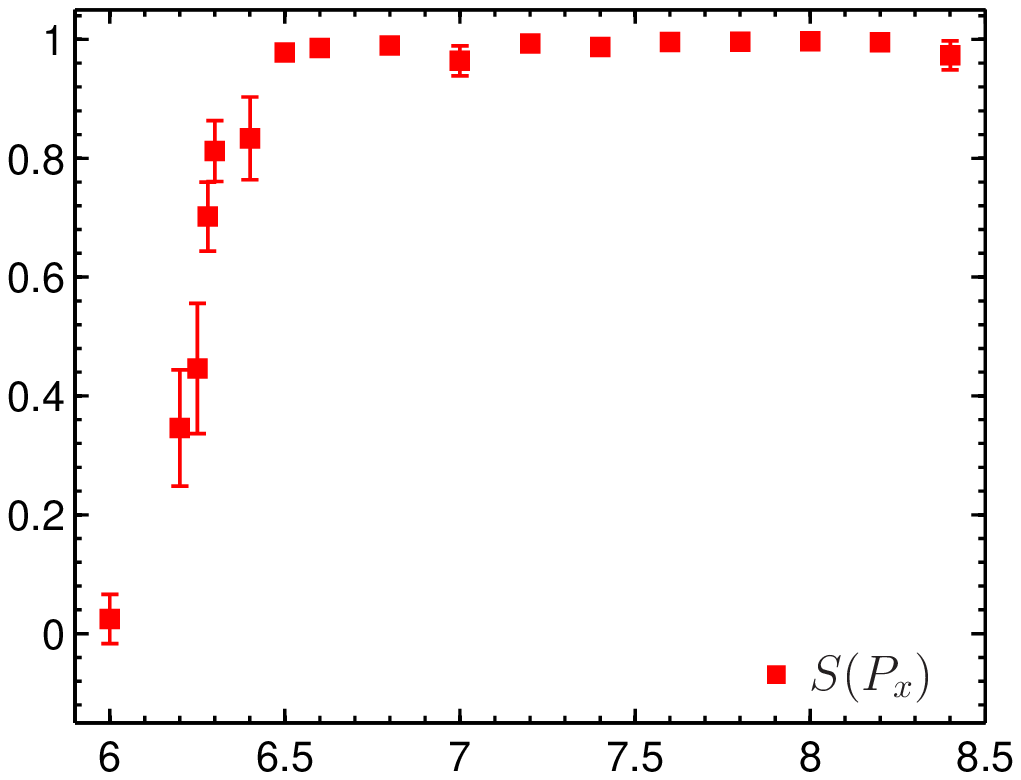, width=.47\textwidth}
{$S(P)$ in the $t$ direction  vs. $\beta$ at $am_q=0.2$ for
 $4\!\times\! 10^2\!\times\! 6$ lattices.
\label{fig:4x10x6m0.2tbeta}}
{$S(P)$ in the $x$ direction  vs. $\beta$ at $am_q=0.2$ for
 $4\!\times\! 10^2\!\times\! 6$ lattices.
\label{fig:4x10x6m0.2xbeta}}

\subsection{Connections to pure gauge theory, and a qualitative explanation}

We are unaware of other simulations of QCD-like theories (with dynamical fermions)
that exhibits behavior like the one described above.
However, gauge theories in three, four and five dimensions actually behave
in a similar way.

We give an illustration, using our own simulations. Take a pure gauge theory
with the Wilson gauge action, periodic  in
all four directions. When one direction is short compared to the other ones
we have the familiar situation of a field theory at finite temperature, which
undergoes a confinement-deconfinement transition at a critical $T$
(or equivalently, at a critical value of the bare gauge coupling constant).
For the Wilson action, this critical coupling is about $\beta=5.69$ for $N=4$ and
5.9 for $N=6$.

Now perform simulations with two short but unequal directions. At a low value of $\beta$,
typically close to, but shifted higher from the critical coupling for one short direction,
the Polyakov loop in the short direction will order. If we then decrease the lattice spacing,
we find that the critical coupling at which the Polyakov loop in the next-smallest direction
also orders is pushed to much higher $\beta$ than it would be in the symmetric (one short direction)
case. 

We illustrate this result from simulations with a $(4\times 6\times 12\times 12)$ lattice:
The phase of the Polyakov loop in the $N=6$ direction shows no evidence
of a transition below at least $\beta=6.3$ (see Fig. \ref{fig:puregaugeS}).

Similar behavior has been reported in two contexts.  The authors of Ref. \cite{Farakos:2002zb}
carried out simulations in four and five dimensional $SU(2)$ gauge theory with two
short directions. Their physical motivation was to study beyond-Standard Model scenarios
with gauge fields in the bulk of compact extra dimensions. Their simulations with
$(2\times 4 \times 16\times 16)$ lattices (and five dimensional analogs), 
show what we have just described.

\DOUBLEFIGURE[!t]
{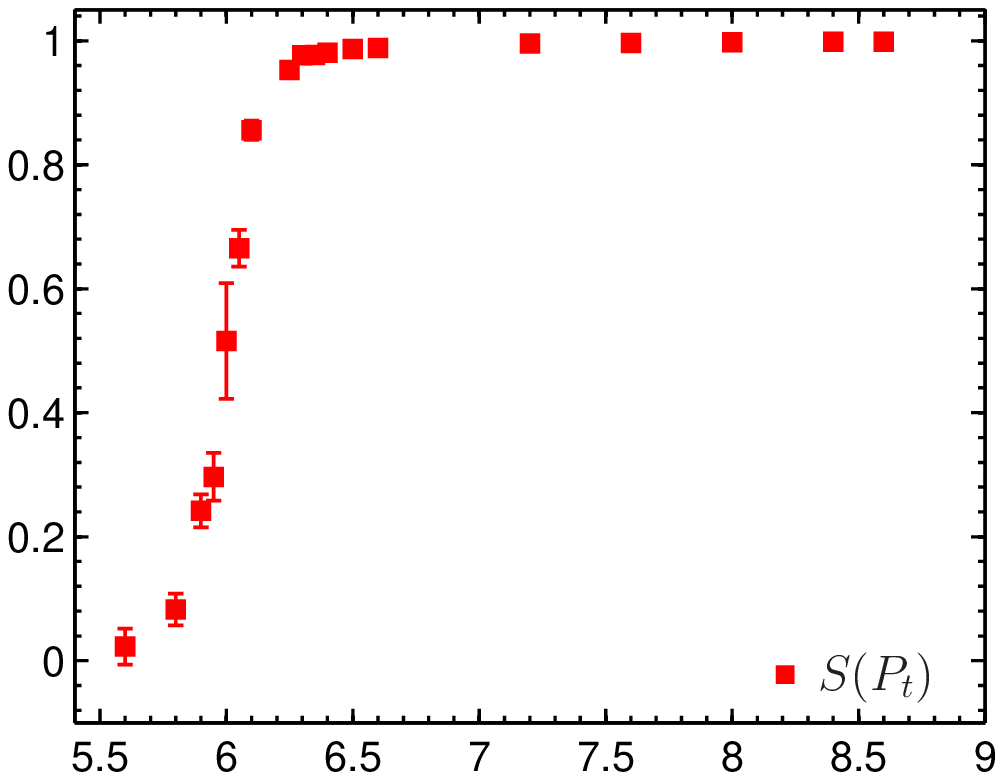, width=.47\textwidth}
{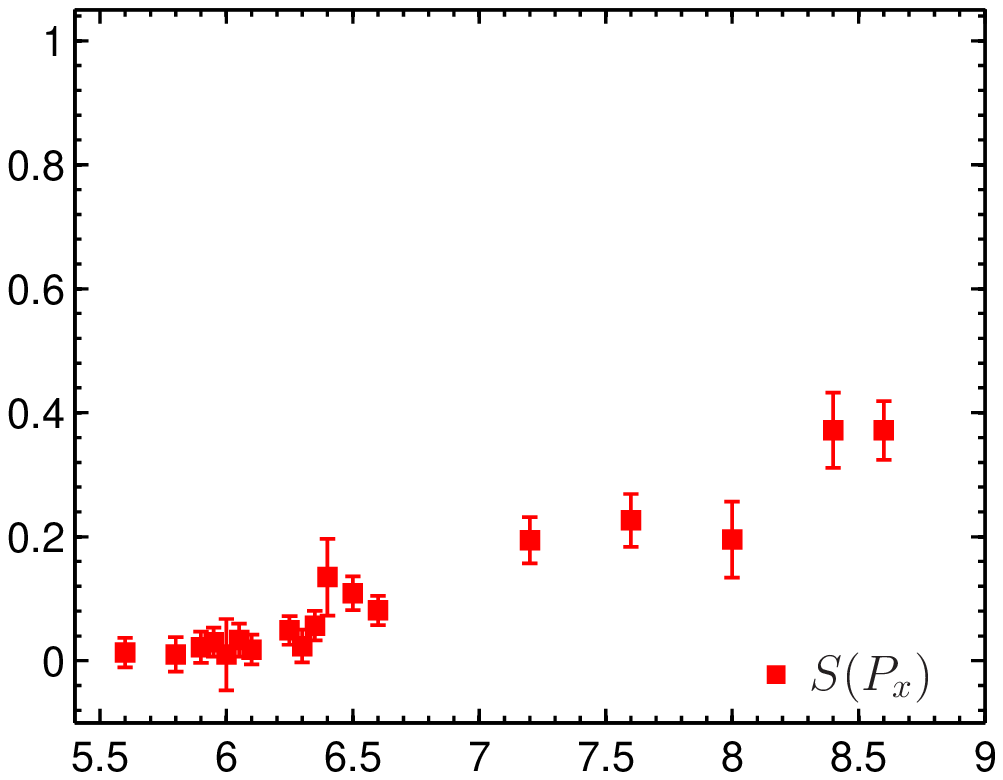, width=.47\textwidth}
{$S(P)$ in the $t$ direction  vs. $\beta$ at $am_q=0.2$ for 
$6\!\times\! 10^2\!\times\! 4$ lattices.
\label{fig:6x10x4m0.2tbeta}}
{$S(P)$ in the $x$ direction vs. $\beta$ at $am_q=0.2$ for
 $6\!\times\! 10^2\!\times\! 4$ lattices.
\label{fig:6x10x4m0.2xbeta}}

The other context is the large-$N$ limit. Bursa and Teper\cite{Bursa:2005tk}
and Narayanan, Neuberger, and Reynoso\cite{Narayanan:2007ug}
have performed simulations of three dimensional gauge theories in asymmetric lattices.
Both groups observe that a sufficiently short length in the shortest direction pushes
the ordering transition in the next-shortest direction to higher $\beta$.

Bursa and Teper describe the phenomenon in terms of 
dimensional reduction. As the shortest direction of the simulation volume is reduced,
the four dimensional gauge theory reduces to a three dimensional gauge theory
with adjoint scalars which are the remnants of the gauge fields oriented
in the short direction \cite{Kajantie:1997tt,Laine:2005ai}.
Bursa and Teper predict that the location of the second transition scales
as $L_2/L_1$ (the ratio of the second-shortest distance to the shortest one),
although their numerical estimate does not seem to be reliable for 
three colors
and $L_2/L_1=6/4$.

\EPSFIGURE[!h]
{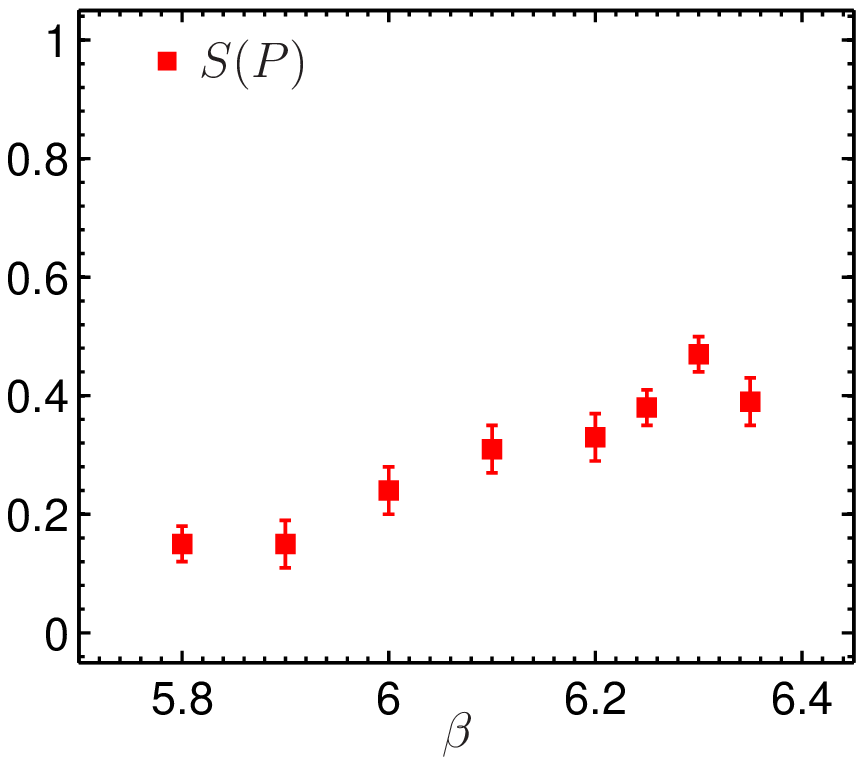, height=.36\textwidth}
{$S(P)$ for Polyakov loops in the $L=6$ direction from pure gauge simulations
 on a $4\!\times\! 6\!\times\! 12\!\times\! 12$ lattice.
\label{fig:puregaugeS}}

Dimensional reduction does give a qualitative explanation for the smooth behavior
seen in Figs. \ref{fig:4x10x6m0.2tbeta},   \ref{fig:6x10x4m0.2xbeta}, and \ref{fig:puregaugeS}:
The gauge group is $SU(3)$. Three dimensional $SU(3)$ pure gauge theory is known to have
a second-order confinement-deconfinement transition with two-dimensional three-state Potts
model exponents\cite{su33d}
as predicted by Svetitsky and Yaffe\cite{Svetitsky:1982gs}.
This already explains why the pure gauge transition is so smooth: it is second order,
probably further smoothed by being on a small lattice. The fermionic results are smoother still.
In four dimensions, the first order nature of the 
 pure gauge transition is robust against the
breaking of $Z(3)$ induced by dynamical fermions.
But no second order transition can survive explicit symmetry breaking, so we can only be
 seeing crossover behavior in this case. This result does not depend on whether
the fermions spontaneously break C, or not.

\section{Conclusion}
We performed simulations of four-flavor, three-color QCD in systems with two
 small spatial directions. These systems can undergo phase transitions in which the
Polyakov loops in different directions can order. When the two short directions
are of equal length, it appears that the Polyakov loops in different directions
are not
strongly correlated, but when one direction is shorter than another one, it
inhibits the ordering in the second-shortest direction.
Dimensional reduction gives a qualitative, though not quantitative, explanation
for the latter phenomenon.
What is amusing about this behavior is that it is shared by pure gauge theories
at both small and large $N$.

\vspace{0.5cm}
\noindent
{\bf Acknowledgments:\\}
T. D. would like to thank B. Svetitsky for discussions, and
R. Narayanan and M. Teper for correspondence.
This work was supported in part by the US Department of Energy.

\end{document}